\documentclass[aps,prl,twocolumn,superscriptaddress,showpacs,amsmath,amssymb]{revtex4}
\usepackage{graphicx} % Include figure files
\usepackage{epstopdf} % Allow to include eps files
\usepackage{dcolumn} % Align table columns on decimal point
\usepackage{xcolor} % Color support
\usepackage{amssymb} % Extended symbol collection
\usepackage{hyperref} % Generate pdfs with hyperlinks
\usepackage[utf8]{inputenc} % Allow UTF8 characters
\usepackage[english]{babel} % All publications are in English
\usepackage[displaymath, mathlines]{lineno} % Line numbers for drafts
\usepackage{blindtext} % For testing
\usepackage[caption=false]{subfig} % caption package not compatible with revtex4

\graphicspath{{figures/}} % Use figures directory for figures

\addto\captionsenglish{}
\addto\captionsenglish{}

\input belle2sym.tex
\newcommand{\BKnn}{\ensuremath{B^{+}\to K^{+}\nu\bar{\nu}}\xspace}
\newcommand{\BJpsiK}{\ensuremath{B^{+}\to K^{+} \jpsi}\xspace}
\def\tomumu{\ensuremath{\to\mu^+\mu^-}}
\def\tomumuslash{\ensuremath{\to\mu^+\!\!\!\!\!\!\!/\,\,\,\,\mu^-\!\!\!\!\!\!\!/\,\,\,\,}}
\newcommand{\BJpsiKmu}{\ensuremath{B^{+}\to K^{+} \jpsi_{\tomumu}}\xspace}
\newcommand{\BJpsiKmuSlash}{\ensuremath{B^{+}\to K^{+} \jpsi_{\tomumuslash}}\xspace}

\newcommand{\pt}{\ensuremath{p_\mathrm{T}}\xspace}
\newcommand{\ptK}{\ensuremath{p_\mathrm{T}(K^{+})}\xspace}
\newcommand{\dz}{\ensuremath{d_{z}}\xspace}
\newcommand{\dr}{\ensuremath{d_{r}}\xspace}

\def\BDT#1{\ensuremath{\mathrm{BDT}_#1}\xspace}

\def\limExp{\ensuremath{2.3 \times 10^{-5}}\xspace}
\def\limObs{\ensuremath{4.1 \times 10^{-5}}\xspace}

\def\CLs{\ensuremath{\mathrm{CL}_s}\xspace}

\begin{document}

% ADD PACS NUMBERS:

\title{% WRITE THE TITLE IN THIS FILE
Search for \BKnn Decays Using an Inclusive Tagging Method at Belle II
}
%%% Paper:    B+ to K+ nu nu-bar
%%% Journal:  Physical Review Letters
%%% Contacts: F. Dattola, S. Glazov, S. Kurz, C. Praz, S. Stefkova
%%% ====================================================================
%%%\newcommand{\instSinica}{Academia Sinica, Taipei 11529, Taiwan}
\newcommand{\instCPPM}{Aix Marseille Universit\'{e}, CNRS/IN2P3, CPPM, 13288 Marseille, France}
\newcommand{\instBeihang}{Beihang University, Beijing 100191, China}
\newcommand{\instBNL}{Brookhaven National Laboratory, Upton, New York 11973, U.S.A.}
\newcommand{\instBINP}{Budker Institute of Nuclear Physics SB RAS, Novosibirsk 630090, Russian Federation}
\newcommand{\instCMU}{Carnegie Mellon University, Pittsburgh, Pennsylvania 15213, U.S.A.}
\newcommand{\instCinvestavIPN}{Centro de Investigacion y de Estudios Avanzados del Instituto Politecnico Nacional, Mexico City 07360, Mexico}
\newcommand{\instPrague}{Faculty of Mathematics and Physics, Charles University, 121 16 Prague, Czech Republic}
\newcommand{\instChiangMai}{Chiang Mai University, Chiang Mai 50202, Thailand}
\newcommand{\instChiba}{Chiba University, Chiba 263-8522, Japan}
\newcommand{\instChonnam}{Chonnam National University, Gwangju 61186, South Korea}
\newcommand{\instConacyt}{Consejo Nacional de Ciencia y Tecnolog\'{\i}a, Mexico City 03940, Mexico}
\newcommand{\instDESY}{Deutsches Elektronen--Synchrotron, 22607 Hamburg, Germany}
\newcommand{\instDuke}{Duke University, Durham, North Carolina 27708, U.S.A.}
\newcommand{\instITAR}{Institute of Theoretical and Applied Research (ITAR), Duy Tan University, Hanoi 100000, Vietnam}
\newcommand{\instRomaENEA}{ENEA Casaccia, I-00123 Roma, Italy}
\newcommand{\instEri}{Earthquake Research Institute, University of Tokyo, Tokyo 113-0032, Japan}
\newcommand{\instJuelich}{Forschungszentrum J\"{u}lich, 52425 J\"{u}lich, Germany}
\newcommand{\instFuJen}{Department of Physics, Fu Jen Catholic University, Taipei 24205, Taiwan}
\newcommand{\instFudan}{Key Laboratory of Nuclear Physics and Ion-beam Application (MOE) and Institute of Modern Physics, Fudan University, Shanghai 200443, China}
\newcommand{\instGoettingen}{II. Physikalisches Institut, Georg-August-Universit\"{a}t G\"{o}ttingen, 37073 G\"{o}ttingen, Germany}
\newcommand{\instGifu}{Gifu University, Gifu 501-1193, Japan}
\newcommand{\instSOKENDAI}{The Graduate University for Advanced Studies (SOKENDAI), Hayama 240-0193, Japan}
\newcommand{\instGyeongsang}{Gyeongsang National University, Jinju 52828, South Korea}
\newcommand{\instHanyang}{Department of Physics and Institute of Natural Sciences, Hanyang University, Seoul 04763, South Korea}
\newcommand{\instKEK}{High Energy Accelerator Research Organization (KEK), Tsukuba 305-0801, Japan}
\newcommand{\instJPARC}{J-PARC Branch, KEK Theory Center, High Energy Accelerator Research Organization (KEK), Tsukuba 305-0801, Japan}
\newcommand{\instHSE}{National Research University Higher School of Economics, Moscow 101000, Russian Federation}
\newcommand{\instIISER}{Indian Institute of Science Education and Research Mohali, SAS Nagar, 140306, India}
\newcommand{\instIITBhubaneswar}{Indian Institute of Technology Bhubaneswar, Satya Nagar 751007, India}
\newcommand{\instIITGuwahati}{Indian Institute of Technology Guwahati, Assam 781039, India}
\newcommand{\instIITHyderabad}{Indian Institute of Technology Hyderabad, Telangana 502285, India}
\newcommand{\instIITMadras}{Indian Institute of Technology Madras, Chennai 600036, India}
\newcommand{\instIndiana}{Indiana University, Bloomington, Indiana 47408, U.S.A.}
\newcommand{\instIHEPRussia}{Institute for High Energy Physics, Protvino 142281, Russian Federation}
\newcommand{\instHEPHYVienna}{Institute of High Energy Physics, Vienna 1050, Austria}
\newcommand{\instHiroshima}{Hiroshima University, Higashi-Hiroshima, Hiroshima 739-8530, Japan}
\newcommand{\instIHEPChina}{Institute of High Energy Physics, Chinese Academy of Sciences, Beijing 100049, China}
\newcommand{\instIPP}{Institute of Particle Physics (Canada), Victoria, British Columbia V8W 2Y2, Canada}
\newcommand{\instIOP}{Institute of Physics, Vietnam Academy of Science and Technology (VAST), Hanoi, Vietnam}
\newcommand{\instIFIC}{Instituto de Fisica Corpuscular, Paterna 46980, Spain}
\newcommand{\instFrascati}{INFN Laboratori Nazionali di Frascati, I-00044 Frascati, Italy}
\newcommand{\instNapoliINFN}{INFN Sezione di Napoli, I-80126 Napoli, Italy}
\newcommand{\instPadovaINFN}{INFN Sezione di Padova, I-35131 Padova, Italy}
\newcommand{\instPerugiaINFN}{INFN Sezione di Perugia, I-06123 Perugia, Italy}
\newcommand{\instPisaINFN}{INFN Sezione di Pisa, I-56127 Pisa, Italy}
\newcommand{\instRomaINFN}{INFN Sezione di Roma, I-00185 Roma, Italy}
\newcommand{\instRomaTreINFN}{INFN Sezione di Roma Tre, I-00146 Roma, Italy}
\newcommand{\instTorinoINFN}{INFN Sezione di Torino, I-10125 Torino, Italy}
\newcommand{\instTriesteINFN}{INFN Sezione di Trieste, I-34127 Trieste, Italy}
\newcommand{\instJAEA}{Advanced Science Research Center, Japan Atomic Energy Agency, Naka 319-1195, Japan}
\newcommand{\instMainz}{Johannes Gutenberg-Universit\"{a}t Mainz, Institut f\"{u}r Kernphysik, D-55099 Mainz, Germany}
\newcommand{\instGiessen}{Justus-Liebig-Universit\"{a}t Gie\ss{}en, 35392 Gie\ss{}en, Germany}
\newcommand{\instKarlsruhe}{Institut f\"{u}r Experimentelle Teilchenphysik, Karlsruher Institut f\"{u}r Technologie, 76131 Karlsruhe, Germany}
\newcommand{\instISU}{Iowa State University, Ames, Iowa 50011, U.S.A.}
\newcommand{\instKitasato}{Kitasato University, Sagamihara 252-0373, Japan}
\newcommand{\instKISTI}{Korea Institute of Science and Technology Information, Daejeon 34141, South Korea}
\newcommand{\instKoreaUnivKU}{Korea University, Seoul 02841, South Korea}
\newcommand{\instKSU}{Kyoto Sangyo University, Kyoto 603-8555, Japan}
\newcommand{\instKyungpook}{Kyungpook National University, Daegu 41566, South Korea}
\newcommand{\instLPI}{P.N. Lebedev Physical Institute of the Russian Academy of Sciences, Moscow 119991, Russian Federation}
\newcommand{\instLNNU}{Liaoning Normal University, Dalian 116029, China}
\newcommand{\instLMU}{Ludwig Maximilians University, 80539 Munich, Germany}
\newcommand{\instLuther}{Luther College, Decorah, Iowa 52101, U.S.A.}
\newcommand{\instMNITJaipur}{Malaviya National Institute of Technology Jaipur, Jaipur 302017, India}
\newcommand{\instMPP}{Max-Planck-Institut f\"{u}r Physik, 80805 M\"{u}nchen, Germany}
\newcommand{\instMPGHLL}{Semiconductor Laboratory of the Max Planck Society, 81739 M\"{u}nchen, Germany}
\newcommand{\instMcGill}{McGill University, Montr\'{e}al, Qu\'{e}bec, H3A 2T8, Canada}
\newcommand{\instMEPhI}{Moscow Physical Engineering Institute, Moscow 115409, Russian Federation}
\newcommand{\instNagoya}{Graduate School of Science, Nagoya University, Nagoya 464-8602, Japan}
\newcommand{\instNagoyaIAR}{Institute for Advanced Research, Nagoya University, Nagoya 464-8602, Japan}
\newcommand{\instNagoyaKMI}{Kobayashi-Maskawa Institute, Nagoya University, Nagoya 464-8602, Japan}
\newcommand{\instNaraWu}{Nara Women's University, Nara 630-8506, Japan}
\newcommand{\instNTUTaiwan}{Department of Physics, National Taiwan University, Taipei 10617, Taiwan}
\newcommand{\instNUUTaiwan}{National United University, Miao Li 36003, Taiwan}
\newcommand{\instKrakow}{H. Niewodniczanski Institute of Nuclear Physics, Krakow 31-342, Poland}
\newcommand{\instNiigata}{Niigata University, Niigata 950-2181, Japan}
\newcommand{\instNSU}{Novosibirsk State University, Novosibirsk 630090, Russian Federation}
\newcommand{\instOkinawa}{Okinawa Institute of Science and Technology, Okinawa 904-0495, Japan}
\newcommand{\instOsakaCity}{Osaka City University, Osaka 558-8585, Japan}
\newcommand{\instRCNP}{Research Center for Nuclear Physics, Osaka University, Osaka 567-0047, Japan}
\newcommand{\instPNNL}{Pacific Northwest National Laboratory, Richland, Washington 99352, U.S.A.}
\newcommand{\instPanjab}{Panjab University, Chandigarh 160014, India}
\newcommand{\instPanjabPAU}{Punjab Agricultural University, Ludhiana 141004, India}
\newcommand{\instRIKENMSL}{Meson Science Laboratory, Cluster for Pioneering Research, RIKEN, Saitama 351-0198, Japan}
\newcommand{\instSeoul}{Seoul National University, Seoul 08826, South Korea}
\newcommand{\instSPU}{Showa Pharmaceutical University, Tokyo 194-8543, Japan}
\newcommand{\instSoochow}{Soochow University, Suzhou 215006, China}
\newcommand{\instSoongsil}{Soongsil University, Seoul 06978, South Korea}
\newcommand{\instLjubljanaJSI}{J. Stefan Institute, 1000 Ljubljana, Slovenia}
\newcommand{\instKyiv}{Taras Shevchenko National Univ. of Kiev, Kiev, Ukraine}
\newcommand{\instTata}{Tata Institute of Fundamental Research, Mumbai 400005, India}
\newcommand{\instTUM}{Department of Physics, Technische Universit\"{a}t M\"{u}nchen, 85748 Garching, Germany}
\newcommand{\instTelAviv}{Tel Aviv University, School of Physics and Astronomy, Tel Aviv, 69978, Israel}
\newcommand{\instToho}{Toho University, Funabashi 274-8510, Japan}
\newcommand{\instTohoku}{Department of Physics, Tohoku University, Sendai 980-8578, Japan}
\newcommand{\instTitech}{Tokyo Institute of Technology, Tokyo 152-8550, Japan}
\newcommand{\instTokyoMetropolitan}{Tokyo Metropolitan University, Tokyo 192-0397, Japan}
\newcommand{\instUAS}{Universidad Autonoma de Sinaloa, Sinaloa 80000, Mexico}
\newcommand{\instNapoliUNIV}{Dipartimento di Scienze Fisiche, Universit\`{a} di Napoli Federico II, I-80126 Napoli, Italy}
\newcommand{\instPadovaUNIV}{Dipartimento di Fisica e Astronomia, Universit\`{a} di Padova, I-35131 Padova, Italy}
\newcommand{\instPerugiaUNIV}{Dipartimento di Fisica, Universit\`{a} di Perugia, I-06123 Perugia, Italy}
\newcommand{\instPisaUNIV}{Dipartimento di Fisica, Universit\`{a} di Pisa, I-56127 Pisa, Italy}
\newcommand{\instRomaTreUNIV}{Dipartimento di Matematica e Fisica, Universit\`{a} di Roma Tre, I-00146 Roma, Italy}
\newcommand{\instTorinoUNIV}{Dipartimento di Fisica, Universit\`{a} di Torino, I-10125 Torino, Italy}
\newcommand{\instTriesteUNIV}{Dipartimento di Fisica, Universit\`{a} di Trieste, I-34127 Trieste, Italy}
\newcommand{\instMontreal}{Universit\'{e} de Montr\'{e}al, Physique des Particules, Montr\'{e}al, Qu\'{e}bec, H3C 3J7, Canada}
\newcommand{\instIJCLab}{Universit\'{e} Paris-Saclay, CNRS/IN2P3, IJCLab, 91405 Orsay, France}
\newcommand{\instIPHC}{Universit\'{e} de Strasbourg, CNRS, IPHC, UMR 7178, 67037 Strasbourg, France}
\newcommand{\instAdelaide}{Department of Physics, University of Adelaide, Adelaide, South Australia 5005, Australia}
\newcommand{\instBonn}{University of Bonn, 53115 Bonn, Germany}
\newcommand{\instUBC}{University of British Columbia, Vancouver, British Columbia, V6T 1Z1, Canada}
\newcommand{\instCincinnati}{University of Cincinnati, Cincinnati, Ohio 45221, U.S.A.}
\newcommand{\instFlorida}{University of Florida, Gainesville, Florida 32611, U.S.A.}
\newcommand{\instHawaii}{University of Hawaii, Honolulu, Hawaii 96822, U.S.A.}
\newcommand{\instHeidelberg}{University of Heidelberg, 68131 Mannheim, Germany}
\newcommand{\instLjubljanaUniLJ}{Faculty of Mathematics and Physics, University of Ljubljana, 1000 Ljubljana, Slovenia}
\newcommand{\instLouisville}{University of Louisville, Louisville, Kentucky 40292, U.S.A.}
\newcommand{\instMalaya}{National Centre for Particle Physics, University Malaya, 50603 Kuala Lumpur, Malaysia}
\newcommand{\instLjubljanaUM}{Faculty of Chemistry and Chemical Engineering, University of Maribor, 2000 Maribor, Slovenia}
\newcommand{\instMelbourne}{School of Physics, University of Melbourne, Victoria 3010, Australia}
\newcommand{\instMississippi}{University of Mississippi, University, Mississippi 38677, U.S.A.}
\newcommand{\instUOM}{University of Miyazaki, Miyazaki 889-2192, Japan}
\newcommand{\instPittsburgh}{University of Pittsburgh, Pittsburgh, Pennsylvania 15260, U.S.A.}
\newcommand{\instUSTC}{University of Science and Technology of China, Hefei 230026, China}
\newcommand{\instSAlabama}{University of South Alabama, Mobile, Alabama 36688, U.S.A.}
\newcommand{\instSCarolina}{University of South Carolina, Columbia, South Carolina 29208, U.S.A.}
\newcommand{\instSydney}{School of Physics, University of Sydney, New South Wales 2006, Australia}
\newcommand{\instUTokyo}{Department of Physics, University of Tokyo, Tokyo 113-0033, Japan}
\newcommand{\instIPMU}{Kavli Institute for the Physics and Mathematics of the Universe (WPI), University of Tokyo, Kashiwa 277-8583, Japan}
\newcommand{\instVictoria}{University of Victoria, Victoria, British Columbia, V8W 3P6, Canada}
\newcommand{\instVPI}{Virginia Polytechnic Institute and State University, Blacksburg, Virginia 24061, U.S.A.}
\newcommand{\instWayneState}{Wayne State University, Detroit, Michigan 48202, U.S.A.}
\newcommand{\instYamagata}{Yamagata University, Yamagata 990-8560, Japan}
\newcommand{\instYerevan}{Alikhanyan National Science Laboratory, Yerevan 0036, Armenia}
\newcommand{\instYonsei}{Yonsei University, Seoul 03722, South Korea}
\newcommand{\instZZU}{Zhengzhou University, Zhengzhou 450001, China}
%%%\affiliation{\instSinica}
\affiliation{\instCPPM}
%%%\affiliation{\instBeihang}
%%%\affiliation{\instBUAP}
\affiliation{\instBNL}
\affiliation{\instBINP}
\affiliation{\instCMU}
\affiliation{\instCinvestavIPN}
\affiliation{\instPrague}
\affiliation{\instChiangMai}
\affiliation{\instChiba}
%%%\affiliation{\instChonnam}
%%%\affiliation{\instChula}
\affiliation{\instConacyt}
\affiliation{\instDESY}
\affiliation{\instDuke}
\affiliation{\instITAR}
\affiliation{\instRomaENEA}
\affiliation{\instEri}
\affiliation{\instJuelich}
%%%\affiliation{\instFuJen}
\affiliation{\instFudan}
\affiliation{\instGoettingen}
\affiliation{\instGifu}
\affiliation{\instSOKENDAI}
%%%\affiliation{\instGyeongsang}
\affiliation{\instHanyang}
\affiliation{\instKEK}
\affiliation{\instJPARC}
\affiliation{\instHSE}
\affiliation{\instIISER}
\affiliation{\instIITBhubaneswar}
%%%\affiliation{\instIITGuwahati}
\affiliation{\instIITHyderabad}
\affiliation{\instIITMadras}
\affiliation{\instIndiana}
\affiliation{\instIHEPRussia}
\affiliation{\instHEPHYVienna}
\affiliation{\instHiroshima}
\affiliation{\instIHEPChina}
%%%\affiliation{\instChennai}
\affiliation{\instIPP}
\affiliation{\instIOP}
\affiliation{\instIFIC}
\affiliation{\instFrascati}
\affiliation{\instNapoliINFN}
\affiliation{\instPadovaINFN}
\affiliation{\instPerugiaINFN}
\affiliation{\instPisaINFN}
\affiliation{\instRomaINFN}
\affiliation{\instRomaTreINFN}
\affiliation{\instTorinoINFN}
\affiliation{\instTriesteINFN}
\affiliation{\instJAEA}
\affiliation{\instMainz}
\affiliation{\instGiessen}
\affiliation{\instKarlsruhe}
\affiliation{\instISU}
%%%\affiliation{\instKennesaw}
\affiliation{\instKitasato}
\affiliation{\instKISTI}
\affiliation{\instKoreaUnivKU}
%%%\affiliation{\instKSU}
%%%\affiliation{\instKyotoU}
\affiliation{\instKyungpook}
\affiliation{\instLPI}
\affiliation{\instLNNU}
\affiliation{\instLMU}
\affiliation{\instLuther}
\affiliation{\instMNITJaipur}
\affiliation{\instMPP}
%%%\affiliation{\instMPGHLL}
\affiliation{\instMcGill}
%%%\affiliation{\instMETU}
\affiliation{\instMEPhI}
\affiliation{\instNagoya}
\affiliation{\instNagoyaKMI}
%%%\affiliation{\instNNU}
\affiliation{\instNaraWu}
%%%\affiliation{\instUNAM}
\affiliation{\instNTUTaiwan}
%%%\affiliation{\instNUUTaiwan}
\affiliation{\instKrakow}
\affiliation{\instNiigata}
\affiliation{\instNSU}
%%%\affiliation{\instOkinawa}
\affiliation{\instOsakaCity}
\affiliation{\instRCNP}
\affiliation{\instPNNL}
\affiliation{\instPanjab}
%%%\affiliation{\instPeking}
\affiliation{\instPanjabPAU}
\affiliation{\instRIKENMSL}
%%%\affiliation{\instRIKEN}
%%%\affiliation{\instXavier}
%%%\affiliation{\instSeoul}
%%%\affiliation{\instShandong}
\affiliation{\instSPU}
\affiliation{\instSoochow}
\affiliation{\instSoongsil}
\affiliation{\instLjubljanaJSI}
\affiliation{\instKyiv}
\affiliation{\instTata}
\affiliation{\instTUM}
%%%\affiliation{\instECUTUM}
\affiliation{\instTelAviv}
\affiliation{\instToho}
\affiliation{\instTohoku}
\affiliation{\instTitech}
%%%\affiliation{\instTokyoMetropolitan}
\affiliation{\instUAS}
%%%\affiliation{\instNapoliUNIVA}
\affiliation{\instNapoliUNIV}
\affiliation{\instPadovaUNIV}
\affiliation{\instPerugiaUNIV}
\affiliation{\instPisaUNIV}
%%%\affiliation{\instRomaUNIV}
\affiliation{\instRomaTreUNIV}
\affiliation{\instTorinoUNIV}
\affiliation{\instTriesteUNIV}
\affiliation{\instMontreal}
\affiliation{\instIJCLab}
\affiliation{\instIPHC}
\affiliation{\instAdelaide}
\affiliation{\instBonn}
\affiliation{\instUBC}
\affiliation{\instCincinnati}
\affiliation{\instFlorida}
%%%\affiliation{\instHamburg}
\affiliation{\instHawaii}
%%%\affiliation{\instHeidelberg}
\affiliation{\instLjubljanaUniLJ}
\affiliation{\instLouisville}
\affiliation{\instMalaya}
\affiliation{\instLjubljanaUM}
\affiliation{\instMelbourne}
\affiliation{\instMississippi}
\affiliation{\instUOM}
%%%\affiliation{\instNovaGorica}
\affiliation{\instPittsburgh}
\affiliation{\instUSTC}
\affiliation{\instSAlabama}
%%%\affiliation{\instSCarolina}
\affiliation{\instSydney}
%%%\affiliation{\instTabuk}
\affiliation{\instUTokyo}
\affiliation{\instIPMU}
\affiliation{\instVictoria}
\affiliation{\instVPI}
\affiliation{\instWayneState}
\affiliation{\instYamagata}
\affiliation{\instYerevan}
\affiliation{\instYonsei}
\affiliation{\instZZU}
  \author{F.~Abudin{\'e}n}\affiliation{\instTriesteINFN} % 2250
  \author{I.~Adachi}\affiliation{\instKEK}\affiliation{\instSOKENDAI} % 2590
% \author{R.~Adak}\affiliation{\instFudan} % 6743
  \author{K.~Adamczyk}\affiliation{\instKrakow} % 2239
  \author{P.~Ahlburg}\affiliation{\instBonn} % 2367
% \author{J.~K.~Ahn}\affiliation{\instKoreaUnivKU} % 7423
  \author{H.~Aihara}\affiliation{\instUTokyo} % 2223
  \author{N.~Akopov}\affiliation{\instYerevan} % 9443
  \author{A.~Aloisio}\affiliation{\instNapoliUNIV}\affiliation{\instNapoliINFN} % 2194
% \author{F.~Ameli}\affiliation{\instRomaINFN} % 4683
% \author{L.~Andricek}\affiliation{\instMPGHLL} % 2098
  \author{N.~Anh~Ky}\affiliation{\instIOP}\affiliation{\instITAR} % 2218
  \author{D.~M.~Asner}\affiliation{\instBNL} % 4684
  \author{H.~Atmacan}\affiliation{\instCincinnati} % 2538
% \author{V.~Aulchenko}\affiliation{\instBINP}\affiliation{\instNSU} % 8183
% ...
% 2387
% \author{V.~Zhulanov}\affiliation{\instBINP}\affiliation{\instNSU} % 4983
% \author{A.~Zupanc}\affiliation{\instLjubljanaJSI} % 2543
  \collaboration{Belle II Collaboration}

\begin{abstract}
A search for the flavor-changing neutral-current decay \BKnn is performed at the \belletwo experiment at the SuperKEKB asymmetric energy electron-positron collider.
The data sample corresponds to an integrated luminosity of $63\invfb$ collected at the \Y4S resonance and a sample of $9\invfb$ collected at an energy 60\mev below the resonance. 
Because the measurable decay signature involves only a single charged kaon, a novel measurement approach is used that exploits not only the properties of the \BKnn decay, but also the inclusive properties of the other \B meson in the $\Y4S \to \B\bar{\B}$ event, to suppress the background from other \B meson decays and light-quark pair production. 
This inclusive tagging approach offers a higher signal efficiency compared to previous searches.
No significant signal is observed.
An upper limit on the branching fraction of \BKnn of \limObs is set at the 90\% confidence level.
\end{abstract}

\maketitle

{
Flavor-changing neutral-current transitions, such as $b\to s \nu\bar{\nu}$, are suppressed in the standard model (SM) by the extended Glashow–Iliopoulos–Maiani mechanism \cite{Glashow:1970gm}.
These transitions can only occur at higher orders in SM perturbation theory via weak amplitudes involving the exchange of at least two gauge bosons, as illustrated in \autoref{fig:feynman}.
The absence of charged leptons in the final state reduces the
theoretical uncertainty compared to similar $b\to s \ell\ell$
transitions, which are affected by the breakdown of factorization due to
photon exchange
\cite{Buras:2014fpa}. 
The branching fraction of the \BKnn decay \cite{chargeConj}, which involves a $b\to s \nu\bar{\nu}$ transition, is predicted to be $\left(4.6 \pm 0.5 \right)\times 10^{-6}$, where the main contribution to the uncertainty arises from the $B^{+}\to K^{+}$ transition form factor \cite{Blake:2016olu}.

\begin{figure}[htp]
\centering
\hspace*{\fill}
\subfloat[Penguin diagram]{\includegraphics[width=.48\linewidth]{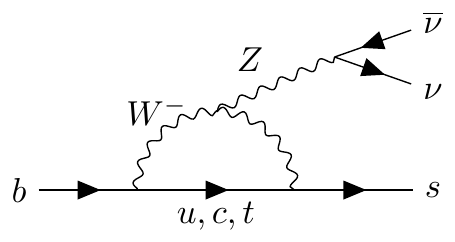}}
\hspace*{\fill}
\hspace*{\fill}
\subfloat[Box diagram]{\includegraphics[width=.48\linewidth]{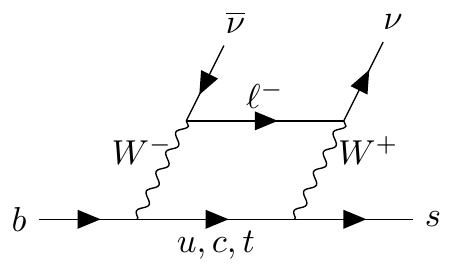}}
\hspace*{\fill}
\caption{The lowest-order quark-level diagrams for the $b\to s \nu\bar{\nu}$ transition in the SM are either of the penguin (a) or the box (b) type.}
\label{fig:feynman}
\end{figure}

Studies of this rare decay are currently of particular interest, as this process offers a complementary probe of potential non-SM physics scenarios that are proposed to explain the tensions with the SM predictions in $b\to s\ellell$ transitions \cite{Descotes-Genon:2020buf} observed in Refs.~\cite{PhysRevLett.122.191801,Aaij_2017,PhysRevLett.111.191801,Aaij_2014,Aaij_2016,Aaij_2020}.
More generally, measurements of the \BKnn decay help constrain models that predict new particles, such as leptoquarks \cite{Becirevic:2018afm}, axions \cite{MartinCamalich:2020dfe}, or dark matter particles \cite{Filimonova:2019tuy}.

The study of the \BKnn decay is experimentally challenging as the final state contains two neutrinos, which leave no signature in the detector and cannot be used to derive information about the signal \B meson. 
Previous searches used \textit{tagged} approaches, where the second \B meson produced in the $e^+e^- \to \Y4S \to \B\bar{\B}$ event is explicitly reconstructed in a hadronic decay \cite{PhysRevLett.86.2950,PhysRevD.87.111103,PhysRevD.87.112005} or in a semileptonic decay \cite{PhysRevD.82.112002,PhysRevD.96.091101}.
This tagging suppresses background events but results in a low signal reconstruction efficiency, typically well below 1\%.
In all analyses reported to date, no evidence for a signal is found, and the current experimental upper limit on the branching fraction is estimated to be $1.6\times10^{-5}$ at 90\% confidence level \cite{Zyla:2020zbs}.

In this search, a novel and independent \textit{inclusive} tagging approach is used, inspired by Ref.~\cite{PhysRevD.101.032007}.
This approach has the benefit of a larger signal efficiency of about 4\%, at the cost of higher background levels.
The method exploits the distinctive topological and kinematic features of the \BKnn decay that distinguish this process from the seven dominant background categories.
These are other decays of charged \B mesons, decays of neutral \B mesons, and the five continuum categories $e^+e^- \to q\bar{q}$ with $q=u,d,s,c$ quarks and $e^+e^- \to \tau^{+} \tau^{-}$.
The signal candidates are reconstructed as a single charged-particle trajectory (track) generated by the kaon, typically carrying higher momentum than background particles.
The remaining tracks and energy deposits, referred to as the "rest of the event" (ROE), can thus be associated with the decay of the accompanying \B meson.
Furthermore, the neutrinos produced in the signal \B meson decay typically carry a significant fraction of its energy.
The resulting "missing momentum" is defined as the momentum needed to cancel the sum of the three-momenta of all reconstructed tracks and energy deposits in the center-of-mass system of the incoming beams.
The specific properties of signal events are captured in a variety of discriminating variables used as inputs for event classifiers to separate signal from background.

This search uses data from $e^+e^-$ collisions produced in 2019 and 2020 by the SuperKEKB collider \cite{Akai:2018mbz}.
The data, corresponding to an integrated luminosity of 63\invfb \cite{Abudinen:2019osb}, were recorded by the Belle II detector at a center-of-mass energy of $\sqrt{s}=10.58\gev$, which corresponds to the \Y4S resonance, and contain 68 million $B\bar{B}$ pairs \cite{Aubert:2004pwa}.
An additional off-resonance sample of 9\invfb integrated luminosity, collected at an energy 60\mev lower than the \Y4S resonance, is used to constrain the yields of continuum events.

The signal and background samples are generated using a variety of event generators, summarized in the Supplemental Material \cite{suppl}, with the detector response simulated by the Belle II Analysis Software Framework (\textsc{basf2}) \cite{Kuhr:2018lps}, interfaced with \textsc{GEANT4} \cite{Agostinelli:2002hh}.
The simulated \BKnn events are weighted according to the SM form-factor calculations from Ref.~\cite{Buras:2014fpa}.

A full description of the Belle II detector is given in Ref.~\cite{Abe:2010gxa}.
The detector consists of several subdetectors arranged in a cylindrical structure around the beam pipe.
Compared to its predecessor Belle \cite{ABASHIAN2002117}, a pixel detector (PXD) has been added at a minimum radius of 1.4\cm.
This improves the resolution of the impact parameter to about 12\mum in the transverse direction for high-momentum tracks \cite{Kou:2018nap}, which helps to reject background events for this analysis.
The PXD is surrounded by a four-layer double-sided silicon strip detector, referred to as the silicon vertex detector, and a central drift chamber (CDC).
A time-of-propagation counter and an aerogel ring-imaging Cherenkov counter cover the barrel and forward endcap regions of the detector, respectively, and are essential for charged-particle identification (PID).
The electromagnetic calorimeter (ECL) makes up the remaining volume inside a superconducting solenoid, which operates at $1.5\,\rm{T}$.
A dedicated detector to identify \KL mesons and muons is installed in the outermost part of the detector.
The $z$ axis of the laboratory frame is defined as the symmetry axis of the solenoid, and the positive direction is approximately given by the incoming electron beam.
The polar angle $\theta$, as well as the longitudinal and the transverse direction are defined with respect to the $z$ axis.
The relevant online event-selection systems (triggers) for this analysis are based either on the number of tracks in the CDC or on the energy deposits in the calorimeter.

The events are reconstructed using \textsc{basf2}.
The trajectories of charged particles are determined using the algorithms described in Ref.~\cite{BERTACCHI2021107610}.
Charged particles are required to have a transverse momentum $\pt>0.1\gevc$, to be within the CDC acceptance ($17^\circ<\theta<150^\circ$), and to have longitudinal and transverse impact parameters with respect to the average interaction point of $|\dz|<3$ and $\dr<0.5\cm$, respectively.
Photons are identified as energy deposits in the ECL exceeding $0.1\gev$ that are within the CDC acceptance and are not matched to tracks.
Each of the charged particles and photons is required to have an energy of less than 5.5\gev to reject misreconstructed objects and cosmic muons.

Events are required to contain no more than ten reconstructed tracks to suppress background events with only a small loss of signal eﬃciency.
Low track-multiplicity background events are suppressed by demanding at least four tracks in the event.
To further suppress such background with a negligible loss of signal events, the total energy from all reconstructed objects in the event must exceed 4\gev and the polar angle $\theta$ of the missing momentum must be between $17^\circ$ and $160^\circ$.

The charged particle with the highest transverse momentum in each event, reconstructed with at least one hit in the PXD, is chosen to be the signal kaon candidate.
Studies on simulated signal events show that the chosen candidate is the signal kaon in 78\% of the cases.
Furthermore, the signal candidate is required to satisfy PID requirements that suppress pion background.
The PID requirements retain 62\% of kaons while removing 97\% of pions for events from the signal region, which is defined below.
Simulated events are weighted to correct the dependence of the efficiency of this selection on the transverse momentum and the polar angle
of the signal candidate, according to the efficiency observed in data.
The remaining charged particles are fit to a common vertex and are attributed, together with the photons, to the ROE.

Simulated signal and background events are used to train binary event classifiers, which are based on the FastBDT algorithm \cite{FastBDTBelleII}, a multivariate method that uses boosted decision trees (BDTs).
Several inputs are considered for this process, including general event-shape variables described in Ref.~\cite{Bevan:2014iga}, as well as variables characterizing the kaon-candidate and the kinematic properties of the ROE.
Moreover, vertices of two and three charged particles, including the kaon candidate, are reconstructed to identify potential kaons from \Dz and \Dp meson decays, and variables describing the fit quality and kinematic properties of the vertices are derived.
Variables that are not well described by the simulation and those that do not contribute to the separation power of the classification are removed.
This results in a set of 51 training variables, summarized in the Supplemental Material \cite{suppl}.

A first binary classifier, \BDT1, is trained on approximately $10^6$ simulated events of each of the seven considered background categories and on the same number of signal events.
The most discriminating variables are found to be event-shape variables, specifically the reduced Fox-Wolfram moment $R_1$, which measures the momentum imbalance in the event where the signal tends to be imbalanced due to signal neutrinos \cite{Fox:1978vw}, and the modified Fox-Wolfram variables that are functions of the missing momentum and of the momentum of the signal kaon candidate \cite{PhysRevLett.91.261801}.

To improve the training performance at high \BDT1 values, a second classifier \BDT2 is trained with the same set of input variables as \BDT1 on events with \BDT1$>0.9$, which corresponds to a signal efficiency of 28\% and a purity of 0.02\%.
The training is performed using a simulated background sample of 100\invfb equivalent luminosity (corresponding to a total of $5\times10^6$ events with $\BDT1>0.9$) and a sample of $1.5\times 10^6$ signal events with $\BDT1>0.9$.
An increase of 35\% in signal purity is achieved by the additional application of \BDT2 on top of \BDT1, when comparing the performance at a signal efficiency of 4\%.
\BDT1 and \BDT2 use the same set of FastBDT parameters \cite{FastBDTBelleII}, which are optimized based on a grid search in the parameter space and are specified in the Supplemental Material \cite{suppl}.

An additional binary classifier is used to correct for mismodeling of continuum simulation, following a data-driven method presented in Ref.~\cite{Martschei_2012}.
More information about the implementation is included in the Supplemental Material \cite{suppl}.
A comparison of simulated continuum events with off-resonance data shows that the application of the derived event weight improves the modeling of all input variables.

A signal region (SR) is defined to be $\BDT1>0.9$ and $\BDT2>0.95$ and is further divided into $3\times3$ bins in the $\BDT2\times\ptK$ space, where \ptK is the transverse momentum of the kaon candidate.
The bin boundaries, decided by minimizing the expected upper limit on the signal branching fraction, are $[0.95, 0.97, 0.99, 1.0]$ in \BDT2 and $[0.5, 2.0, 2.4, 3.5]\gevc$ in \ptK. 
Furthermore, three control regions are used to help constrain the background yields.
The control region CR1 consists of $1\times3$ bins in the $\BDT2\times\ptK$ space, defined at lower values of $\BDT2\in[0.93,0.95]$ and using the same \ptK bins as the SR.
The two other control regions, CR2 and CR3, consist of off-resonance data with identical \BDT2 and \ptK ranges and bins as in the SR and CR1, respectively.

The expected yields of the SM signal and the backgrounds in the SR are 14 and 844 events, respectively, corresponding to a signal efficiency of $4.3\%$.
In most of these background events, a $K^+$ produced in a $D$ meson decay is selected as the signal kaon candidate.

To enable the study of other, non-SM signal models, the fraction of signal events in the SR is studied as a function of the generated dineutrino invariant mass squared $q^2$.  
The efficiency is $13\%$ for $q^2=0$ and drops to zero for $q^2>16$~\ensuremath{{\mathrm{\,Ge\kern -0.1em V^2\!/}c^4}}.
The full distribution can be found in the Supplemental Material \cite{suppl}.

The performance of the classifiers \BDT1 and \BDT2 on data is tested by selecting events with a moderate BDT output of $0.9<\BDT1<0.99$ and $\BDT2<0.7$ in the \Y4S on-resonance data and corresponding simulation.
The study confirms the accurate modeling of the BDT distributions by the simulation for a sample of events that have similar kinematic properties as signal events, while containing only a negligible contribution from signal.

The decay \BJpsiK with $\jpsi\tomumu$ is used as an independent validation channel, exploiting its large branching fraction and distinctive experimental signature.
These events are selected in data and \BJpsiK simulation by requiring the presence of two muons with an invariant mass within 50\mevcc of the known \jpsi mass \cite{Zyla:2020zbs}.
To suppress background events, the variable $\left|\Delta E\right|=\left|E^*_B-\sqrt{s}/2\right|$ is required to be less than 100\mev and the beam-energy constrained mass $M_{\mathrm{bc}}=\sqrt{s/(4c^4)-{p^*_B}^2/c^2}$ is required to exceed 5.25\gevcc, where $E^*_B$ and $p^*_B$ are given by the energy and the magnitude of the three-momentum of the signal \B meson candidate defined in the center-of-mass system of the incoming beams.
This results in 1720 events being selected in the data sample at an expected background contamination of 5\%.
Each event is then reconsidered as a \BKnn event by ignoring the muons from the \jpsi decay and replacing the momentum of the kaon candidate with the generator-level momentum of the kaon in a randomly selected \BKnn event from simulation.
The same modifications are applied to the data and \BJpsiK simulation.
The results are summarized in \autoref{fig:jpsi_bdt}, where the distributions of the output values of both BDTs are shown.
Good agreement between simulation and data is observed for the selected events before (\BJpsiKmu) and after (\BJpsiKmuSlash) the modifications.
The two-sample Kolmogorov-Smirnov $p$ values \cite{Hodges1958TheSP} for the \BDT1 and \BDT2 distributions of simulation and data, after the modifications, are 7\% and 23\%, respectively.
The ratio of the selection efficiencies $\BDT1>0.9,\,\BDT2>0.95$ in data and simulation is found to be $1.06\pm0.10$.

\begin{figure}[htp]
\centering
\includegraphics[width=\linewidth]{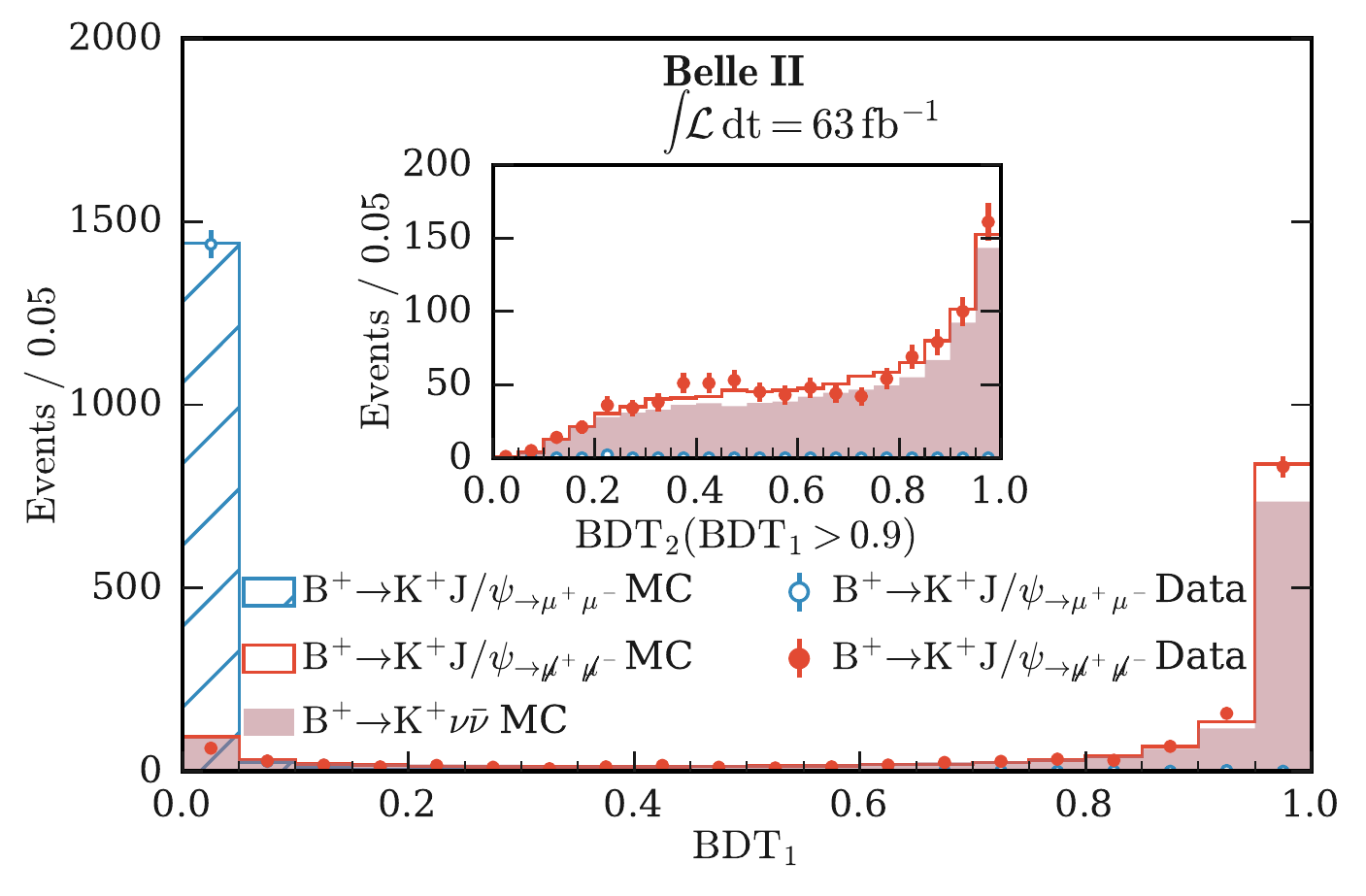}
\caption{
	Distribution of the classifier output \BDT1 (main figure) and \BDT2 for $\BDT1>0.9$ (inset).
	The distributions are shown before (${\jpsi}_{\tomumu}$) and after (${\jpsi}_{\tomumuslash}$) the muon removal and update of the kaon-candidate momentum of selected \BJpsiK events in Monte-Carlo simulation (MC) and data.
	As a reference, the classifier outputs directly obtained from simulated \BKnn signal events are overlaid.
	The simulation histograms are scaled to the total number of \BJpsiK events selected in data.
	}
\label{fig:jpsi_bdt}
\end{figure}

The statistical analysis to determine the signal yields is performed with the \textsc{pyhf} package \cite{pyhf_joss,pyhf}, which constructs a binned likelihood following the \textsc{HistFactory} \cite{Cranmer:1456844} formalism.
The templates for the yields of the signal and the seven background categories are derived from simulation.
The likelihood function is a product of Poisson probability density functions that combine the information from all 24 signal- and control-region bins defined on the on- and off-resonance data.
The systematic uncertainties discussed below are included in the likelihood as nuisance parameters that are event-count modifiers with corresponding constraints modeled as normal probability density functions.
The parameter of interest, the signal strength $\mu$, is defined as a factor relative to the SM expectation and is determined simultaneously with the nuisance parameters using a simultaneous maximum-likelihood fit to the binned distribution of data event counts.    

The leading systematic uncertainty is the normalization uncertainty on the background yields.
The yields of the seven individual background categories are allowed to float independently in the fit.
However, each of them is constrained assuming a normal constraint, centered at the expected background yield obtained from simulation and a standard deviation corresponding to 50\% of the central value.
This value is motivated by a global normalization difference of $(40\pm12)\%$ between the off-resonance data and simulation in the control regions CR2 and CR3 and also covers the uncertainty on the sample luminosity.
The remaining considered systematic uncertainties may also influence the shape of the templates.
Systematic uncertainties originating from the branching fractions of the leading \B meson decays, the PID correction, and the SM form factors are accounted for with three nuisance parameters each to model correlations between the individual SR and CR bins.
The remaining systematic uncertainties arise from the energy miscalibration of hadronic and beam-background calorimeter energy deposits and the tracking inefficiency, and are each accounted for with one nuisance parameter.
The systematic uncertainty due to the limited size of simulated samples is taken into account by one nuisance parameter per bin per background category.
This results in a total of 175 nuisance parameters.

To validate the fitting software, an alternative approach based on a simplified Gaussian likelihood function (\textsc{sghf}) is developed.
Tests of both \textsc{pyhf} and \textsc{sghf} are performed using pseudo-experiments, in which both statistical and systematic uncertainties are taken into account, including background normalizations.
No bias in $\mu$ and its uncertainty is observed, and the $p$ value for the data and fit model compatibility is found to be above 65\%.

Shifts of the nuisance parameters corresponding to the seven background sources are investigated before $\mu$ is revealed.
The parameters corresponding to the continuum background yields are increased by, at most, one standard deviation, which confirms that they are not pulled substantially in the fit given the observed difference in the normalization of the continuum simulation with respect to the off-resonance data.
The background yields in the bins of CR2 and CR3 predicted by the fit are found in agreement with the off-resonance data. 
No shift is observed for the parameters corresponding to the background yields from charged and neutral \B meson decays, which are the dominant contributions in the most sensitive SR bins.

A comparison of the data and fit results in the SR and CR1 is shown in \autoref{fig:yields}.
The corresponding figure for CR2 and CR3 can be found in the Supplemental Material \cite{suppl}.
The signal purity is found to be 6\% in the SR and is as high as 22\% in the three bins with $\BDT2>0.99$.
Continuum events make up 59\% of the background in the SR and 28\% of the events with $\BDT2>0.99$.

\begin{figure}[htp]
\centering
\includegraphics[width=\linewidth]{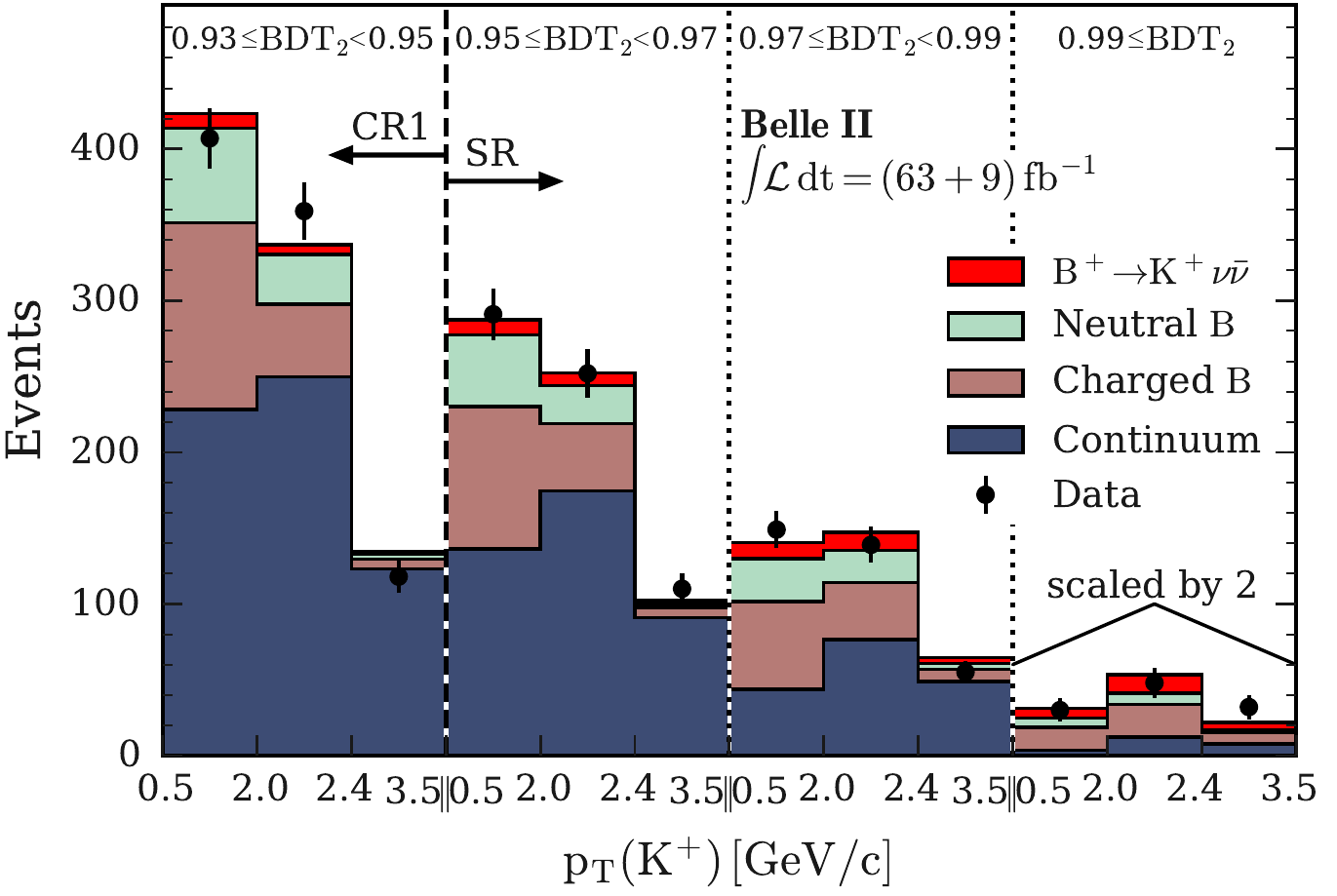}
\caption{
	Yields in on-resonance data and as predicted by the simultaneous fit to the on- and off-resonance data, corresponding to an integrated luminosity of 63 and 9\invfb, respectively.
	The predicted yields are shown individually for charged and neutral \B meson decays and the sum of the five continuum categories.
	The leftmost three bins belong to CR1 with $\BDT2\in[0.93,0.95]$ and the other nine bins correspond to the SR, three for each range of $\BDT2\in[0.95, 0.97, 0.99, 1.0]$.
	Each set of three bins is defined by $\ptK\in[0.5, 2.0, 2.4, 3.5]\gevc$.
	All yields in the rightmost three bins are scaled by a factor of 2.
	}
\label{fig:yields}
\end{figure}

The signal strength is determined by the fit to be $\mu = 4.2^{+3.4}_{-3.2} = 4.2^{+2.9}_{-2.8}(\mathrm{stat}){}^{+1.8}_{-1.6}(\mathrm{syst})$, where the statistical uncertainty is estimated using pseudo-experiments based on Poisson statistics.
The total uncertainty is obtained by a profile likelihood scan, fitting the model with fixed values of $\mu$ around the best-fit value while keeping the other fit parameters free.
The systematic uncertainty is calculated by subtracting the statistical uncertainty in quadrature from the total uncertainty.
An additional 10\% theoretical uncertainty arising from the knowledge of the branching ratio in the SM is not included.
The result corresponds to a branching fraction of the \BKnn decay of $\left[1.9^{+1.6}_{-1.5}\right] \times 10^{-5} = \left[1.9^{+1.3}_{-1.3}(\mathrm{stat}){}^{+0.8}_{-0.7}(\mathrm{syst})\right] \times 10^{-5}$.

This value is statistically compatible with the measurements performed by previous experiments.
Details are given in the Supplemental Material \cite{suppl}.
The uncertainty on the branching fraction is used to define a measure to compare the performance of the individual tagging techniques.
Assuming that this uncertainty scales as the inverse square root of the integrated luminosity
\footnote{
The product $\sigma_{\mathrm{BR}}\sqrt{L}$ of the total branching-fraction uncertainty $\sigma_{\mathrm{BR}}$ and the integrated luminosity of the data sample $L$ is used as a measure, assuming that $\sigma_{\mathrm{BR}}$ scales as $1/\sqrt{L}$.
},
the inclusive approach is more than a factor of 3.5 better per integrated luminosity than the hadronic tagging of Ref.~\cite{PhysRevD.87.111103}, approximately 20\% better than the semileptonic tagging of Ref.~\cite{PhysRevD.96.091101} and approximately 10\% better than the combined hadronic and semileptonic tagging of Ref.~\cite{PhysRevD.87.112005}.
Moreover, the events in the SR differ from the ones selected by the hadronic and semileptonic tagging so that a statistical combination of the measurements provides additional sensitivity.
The inclusive tagging approach can be applied to a variety of decay modes, such as $B^{+}\to K^{*+}\nu\bar{\nu}$ and $B^{+}\to \tau^{+}\nu$, and, because of the increased signal efficiency, this can be done on smaller data samples than required for semileptonic or hadronic tagging.

As no significant signal is observed, the expected and observed upper limits on the branching fraction are determined using the \CLs method \cite{Read_2002}, a modified frequentist approach that is based on a profile likelihood ratio \cite{Cowan:2010js}.
The expected 90\% confidence level (C.L.) upper limit on the \BKnn branching fraction of \limExp is derived assuming a background-only hypothesis.
The observed upper limit is \limObs at the 90\% C.L.
The full distribution of the determined \CLs values is shown in the Supplemental Material \cite{suppl}.

In summary, a search for the rare decay \BKnn is performed using an inclusive tagging approach, which has not previously been used to study this process. 
This analysis uses data corresponding to an integrated luminosity of 63\invfb collected at the \Y4S resonance by the Belle II detector, as well as an off-resonance sample corresponding to 9\invfb.
No statistically significant signal is observed and an upper limit on the branching fraction of \limObs at the 90\% C.L. is set, assuming a SM signal.
This measurement is competitive with previous results for similar integrated luminosities, demonstrating the capability of the inclusive tagging approach, which is widely applicable and expands the future physics reach of Belle II.

}

We thank the SuperKEKB group for the excellent operation of the
accelerator; the KEK cryogenics group for the efficient
operation of the solenoid; the KEK computer group for
on-site computing support; and the raw-data centers at
BNL, DESY, GridKa, IN2P3, and INFN for off-site computing support.
This work was supported by the following funding sources:
%Armenia
Science Committee of the Republic of Armenia Grant No.~20TTCG-1C010;
%Australia
Australian Research Council and research Grants
No.~DP180102629,
No.~DP170102389,
No.~DP170102204,
No.~DP150103061,
No.~FT130100303,
No.~FT130100018,
and
No.~FT120100745;
%Austria
Austrian Federal Ministry of Education, Science and Research,
Austrian Science Fund No.~P 31361-N36, and
Horizon 2020 ERC Starting Grant No.~947006 ``InterLeptons'';
%Canada
Natural Sciences and Engineering Research Council of Canada, Compute Canada and CANARIE;
%China
Chinese Academy of Sciences and research Grant No.~QYZDJ-SSW-SLH011,
National Natural Science Foundation of China and research Grants
No.~11521505,
No.~11575017,
No.~11675166,
No.~11761141009,
No.~11705209,
and
No.~11975076,
LiaoNing Revitalization Talents Program under Contract No.~XLYC1807135,
Shanghai Municipal Science and Technology Committee under Contract No.~19ZR1403000,
Shanghai Pujiang Program under Grant No.~18PJ1401000,
and the CAS Center for Excellence in Particle Physics (CCEPP);
%Czech Republic
the Ministry of Education, Youth, and Sports of the Czech Republic under Contract No.~LTT17020 and
Charles University Grants No.~SVV 260448 and No.~GAUK 404316;
%EU
European Research Council, Seventh Framework PIEF-GA-2013-622527,
Horizon 2020 ERC-Advanced Grants No.~267104 and No.~884719,
Horizon 2020 ERC-Consolidator Grant No.~819127,
Horizon 2020 Marie Sklodowska-Curie Grant Agreement No.~700525 "NIOBE",
and
Horizon 2020 Marie Sklodowska-Curie RISE project JENNIFER2 Grant Agreement No.~822070 (European grants);
%France
L'Institut National de Physique Nucl\'{e}aire et de Physique des Particules (IN2P3) du CNRS (France);
%Germany
BMBF, DFG, HGF, MPG, and AvH Foundation (Germany);
%India
Department of Atomic Energy under Project Identification No.~RTI 4002 and Department of Science and Technology (India);
%Israel
Israel Science Foundation Grant No.~2476/17,
U.S.-Israel Binational Science Foundation Grant No.~2016113, and
Israel Ministry of Science Grant No.~3-16543;
%Italy
Istituto Nazionale di Fisica Nucleare and the research grants BELLE2;
%Japan
Japan Society for the Promotion of Science, Grant-in-Aid for Scientific Research Grants
No.~16H03968,
No.~16H03993,
No.~16H06492,
No.~16K05323,
No.~17H01133,
No.~17H05405,
No.~18K03621,
No.~18H03710,
No.~18H05226,
No.~19H00682, % Niigata
No.~26220706,
and
No.~26400255,
the National Institute of Informatics, and Science Information NETwork 5 (SINET5), 
and
the Ministry of Education, Culture, Sports, Science, and Technology (MEXT) of Japan;  
%Korea
National Research Foundation (NRF) of Korea Grants
No.~2016R1\-D1A1B\-01010135,
No.~2016R1\-D1A1B\-02012900,
No.~2018R1\-A2B\-3003643,
No.~2018R1\-A6A1A\-06024970,
No.~2018R1\-D1A1B\-07047294,
No.~2019K1\-A3A7A\-09033840,
and
No.~2019R1\-I1A3A\-01058933,
Radiation Science Research Institute,
Foreign Large-size Research Facility Application Supporting project,
the Global Science Experimental Data Hub Center of the Korea Institute of Science and Technology Information
and
KREONET/GLORIAD;
%Malaysia
Universiti Malaya RU grant, Akademi Sains Malaysia, and Ministry of Education Malaysia;
%Mexico
% CINVESTAV-IPN, UNAM, UAS, BUAP and CONACYT are funded under
Frontiers of Science Program Contracts
No.~FOINS-296,
No.~CB-221329,
No.~CB-236394,
No.~CB-254409,
and
No.~CB-180023, and No.~SEP-CINVESTAV research Grant No.~237 (Mexico);
%Poland
the Polish Ministry of Science and Higher Education and the National Science Center;
%Russia
the Ministry of Science and Higher Education of the Russian Federation,
Agreement No.~14.W03.31.0026, and
the HSE University Basic Research Program, Moscow;
%Saudi Arabia
University of Tabuk research Grants
No.~S-0256-1438 and No.~S-0280-1439 (Saudi Arabia);
%Slovenia
Slovenian Research Agency and research Grants
No.~J1-9124
and
No.~P1-0135;
%Spain
Agencia Estatal de Investigacion, Spain Grants
No.~FPA2014-55613-P
and
No.~FPA2017-84445-P,
and
No.~CIDEGENT/2018/020 of Generalitat Valenciana;
%Taiwan
Ministry of Science and Technology and research Grants
No.~MOST106-2112-M-002-005-MY3
and
No.~MOST107-2119-M-002-035-MY3,
and the Ministry of Education (Taiwan);
%Thailand
Thailand Center of Excellence in Physics;
%Turkey
TUBITAK ULAKBIM (Turkey);
%Ukraine
Ministry of Education and Science of Ukraine;
%USA
the U.S. National Science Foundation and research Grants
No.~PHY-1807007 % Luther
and
No.~PHY-1913789, % Indiana CEEM
and the U.S. Department of Energy and research Awards
No.~DE-AC06-76RLO1830, % PNNL
No.~DE-SC0007983, % Wayne State
No.~DE-SC0009824, % Florida
No.~DE-SC0009973, % VPI
No.~DE-SC0010073, % South Carolina
No.~DE-SC0010118, % Carnegie Mellon
No.~DE-SC0010504, % Hawaii
No.~DE-SC0011784, % Cincinnati
No.~DE-SC0012704, % BNL
No.~DE-SC0021274; % Mississippi
%last group
and
%Vietnam
the Vietnam Academy of Science and Technology (VAST) under Grant No.~DL0000.05/21-23.

\bibliography{references}

\begin{thebibliography}{47}
\bibitem[{cha()}]{chargeConj}
\bibinfo{note}{Charge-conjugate channels are implied throughout this letter.}
\end{thebibliography}

\end{document}

\makeatletter
\begin{titlepage}
    \centering
    \includegraphics[width=3cm]{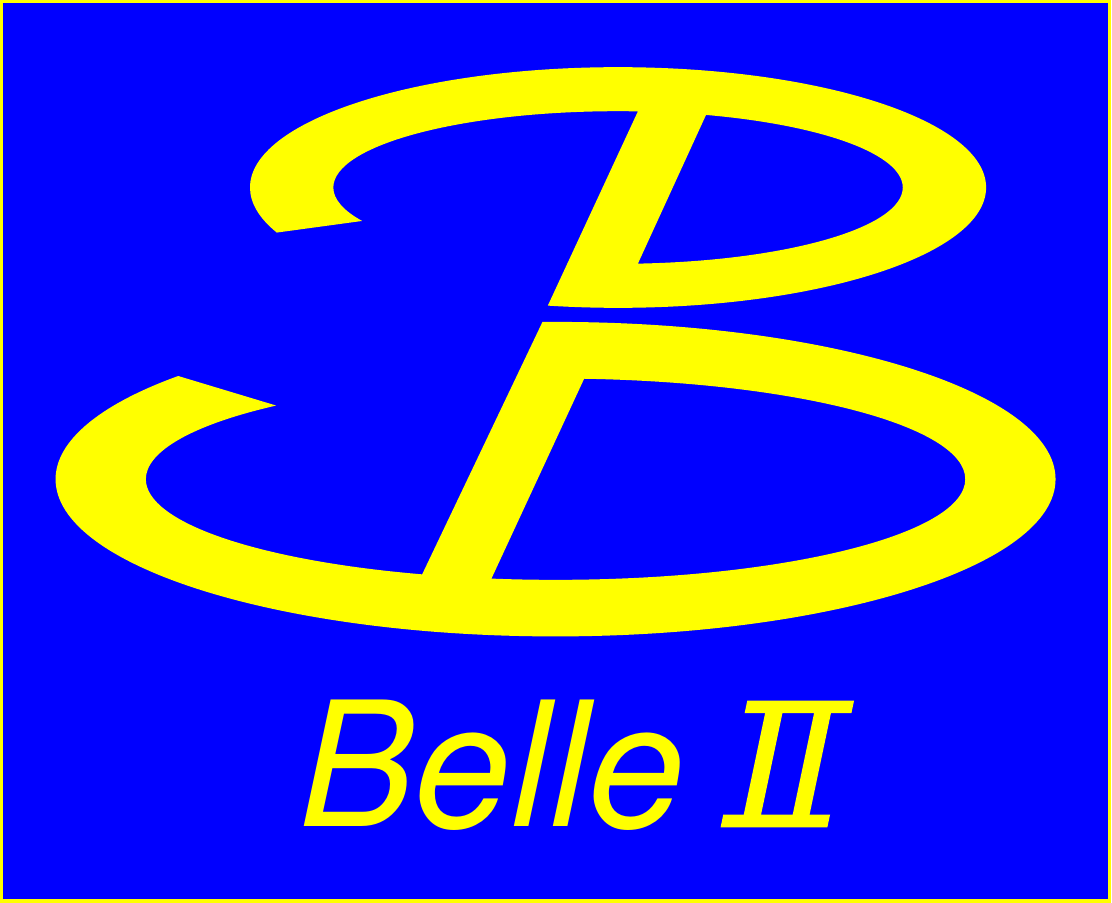}\\[5cm]
    {\Large% WRITE THE TITLE IN THIS FILE
Search for \BKnn Decays Using an Inclusive Tagging Method at Belle II
}\\[.5cm]
    {\large(Supplemental Material)}\\[1cm]
    \today
    \newpage
\end{titlepage}

\section{List of BDT parameters and input variables}

Table \ref{tab:parameters} presents the parameters that are used to train the classifiers (\BDT1, \BDT2).
Furthermore, all input variables are listed below.
Unless otherwise specified, all variables are measured in the laboratory frame.
The variable selection is done by iteratively removing variables from the training and checking the impact of their removal on the binary classification performance, measured with the area under the receiver operating characteristic curve \cite{FAWCETT2006861}.

\begin{table}[htp]
  \caption{\small FastBDT parameter values \cite{FastBDTBelleII}.}
  \label{tab:parameters}
\begin{center}\begin{tabular}{l@{\hskip 1cm}l@{\hskip 1cm}l}
    \hline
    Parameter & Value \\
    \hline
    Number of trees & 2000 \\
    Tree depth & 4 \\
    Shrinkage & 0.1 \\
    Sampling rate & 0.8 \\
    Number of equal-frequency bins & 16 \\
    \hline
  \end{tabular}\end{center}
\end{table}

For a given track, the point of closest approach (POCA) is defined as the point on the track that minimizes the distance to a line $d$ passing through the average interaction point and parallel to the $z$-axis, defined as the symmetry axis of the solenoid.
The transverse impact parameter $\dr$ is defined as this minimal distance and the longitudinal impact parameter $\dz$ is defined as the $z$-coordinate of the POCA with respect to the average interaction point \cite{BERTACCHI2021107610}. \newline

\noindent \underline{Variables related to the kaon candidate}
\begin{itemize}
\item Azimuthal angle of the kaon momentum at the POCA
\item $\dr$ and $\dz$ of the kaon track
\item Cosine of the polar angle of the kaon 3-momentum at the POCA
\end{itemize}
Variables related to the kaon candidate do not include $\ptK$, because the data are binned in this variable and in $\BDT2$ in the last stage of the analysis. \newline

\noindent \underline{Variables related to the tracks and energy deposits of}\\ \underline{the rest of the event (ROE)}
\begin{itemize}
\item Three variables corresponding to the $x$, $y$, $z$ components of the vector from the average interaction point to the ROE vertex
\item $\dr$ and $\dz$ of the kaon track with respect to the ROE vertex
\item Invariant mass of the ROE
\item $\chi^2$ of the ROE vertex fit
\item $p$-value of the ROE vertex fit
\item Variance of the transverse momentum of the ROE tracks
\item Polar angle of the ROE momentum
\item Magnitude of the ROE momentum
\item Magnitude of the ROE thrust in the center-of-mass system (CMS) \cite{Bevan:2014iga}
\item Difference between the ROE energy in the CMS and $\sqrt{s}/2$
\end{itemize}

\noindent \underline{Variables related to the entire event}
\begin{itemize}
\item Magnitude of the event thrust in the CMS
\item Number of charged lepton candidates (\epm or {\ensuremath{\mu^\pm}\xspace})
\item Number of photon candidates, number of charged particle candidates and sum of the two
\item Square of the total charge of tracks
\item Cosine of the polar angle of the thrust axis in the CMS
\item Zeroth-order and second-order harmonic moments with respect to the thrust axis in the CMS \cite{Fox:1978vw}
\item Seven modified Fox-Wolfram moments calculated in the CMS \cite{PhysRevLett.91.261801}
\item Polar angle of the missing three-momentum in the CMS
\item Square of the missing invariant mass
\item Event sphericity in the CMS \cite{Bevan:2014iga}
\item First, second and third normalized Fox-Wolfram moments in the CMS \cite{Fox:1978vw}
\item Cosine of the angle between the kaon track and the ROE thrust axis in the CMS
\end{itemize}

\noindent \underline{Variables related to the $\Dz/\Dp$ suppression}\\ As mentioned in the main text, $\Dz$ candidates are obtained by fitting the kaon candidate track and each track of opposite charge in the ROE to a common vertex; $\Dp$ candidates are obtained by fitting the kaon candidate track and two ROE tracks of appropriate charges. In both cases, the best candidate is the one having the best vertex fit quality.
\begin{itemize}
\item $\dr$ and $\dz$ of the best $\Dz$ candidate vertex and of the best $\Dp$ candidate vertex
\item $\chi^2$ of the best $\Dz$ candidate vertex fit and the best $\Dp$ candidate vertex fit
\item Mass of the best $\Dz$ candidate
\item Median $p$-value of the vertex fits of the $\Dz$ candidates
\end{itemize}

\section{Event Generators}

A variety of event generators are used to simulate the considered signal and background samples.
The decays of charged and neutral \B mesons are simulated using the \texttt{EVTGEN} event generator \cite{Lange:2001uf}.
\texttt{KKMC} \cite{Jadach:1999vf} is used to generate the $q\bar{q}$ pairs, with \texttt{PYTHIA8} \cite{Sjostrand:2014zea} to simulate their hadronization and \texttt{EVTGEN} to model the decays of the generated mesons.
\texttt{KKMC} and \texttt{TAUOLA} \cite{Jadach:1990mz} are employed to simulate $e^+e^- \to \tau^{+} \tau^{-}$ events.
For all samples, the \textit{Belle II Analysis Software Framework} \cite{Kuhr:2018lps}, interfaced with \texttt{GEANT4} \cite{Agostinelli:2002hh}, is used to simulate the detector response.

\section{Continuum Reweighting}

As stated in the main text, mismodeling of continuum simulation is corrected for following a data-driven procedure presented in Ref.~\cite{Martschei_2012}.
A binary classifier, \BDTc, is trained to separate the off-resonance data and continuum simulation using the same set of input variables and parameters as \BDT1 and \BDT2.
\BDTc is trained with events that satisfy \BDT1$>0.9$ in the off-resonance data and an independent 100\invfb sample of simulated continuum events.
If $p\in[0,1]$ is the classifier output for a given continuum event, the ratio $p/(1-p)$ can be interpreted as an estimate of the likelihood ratio $\mathcal{L}(\mathrm{data})/\mathcal{L}(\mathrm{MC})$, where $\mathcal{L}(\mathrm{data})$ $(\mathcal{L}(\mathrm{MC}))$ is the likelihood of the continuum event being from data (simulation).
This event weight is applied to the simulated continuum events after the final selection.
Comparison of simulated continuum events with off-resonance data shows that the application of this weight improves the modeling of the input variables.

\newpage

\section{Supplemental figures}

\begin{figure}[htp]
\centering
\includegraphics[width=\linewidth]{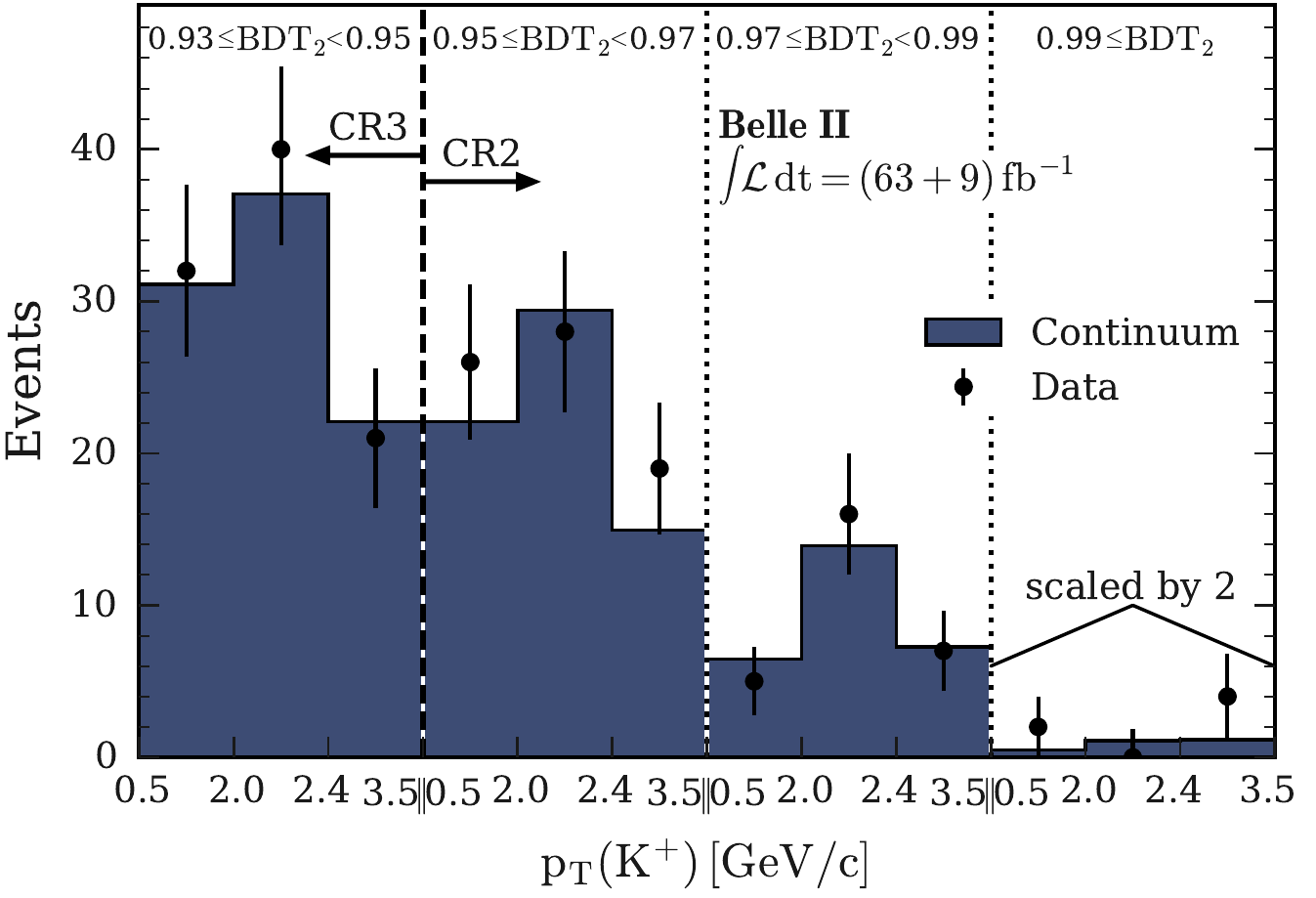}
\caption{
  Yields in off-resonance data and as predicted by the simultaneous fit to the on- and off-resonance data, corresponding to an integrated luminosity of 63\invfb and 9\invfb, respectively.
  The predicted yields of the five continuum categories are summarized.
  The leftmost three bins belong to CR3 with $\BDT2\in[0.93,0.95]$ and the other nine bins correspond to the CR2, three for each range of $\BDT2\in[0.95, 0.97, 0.99, 1.0]$.
  Each set of three bins is defined by $\ptK\in[0.5, 2.0, 2.4, 3.5]\gevc$.
  All yields in the rightmost three bins are scaled by a factor of two.}
\end{figure}

\begin{figure}[htp]
\centering
\includegraphics[width=\linewidth]{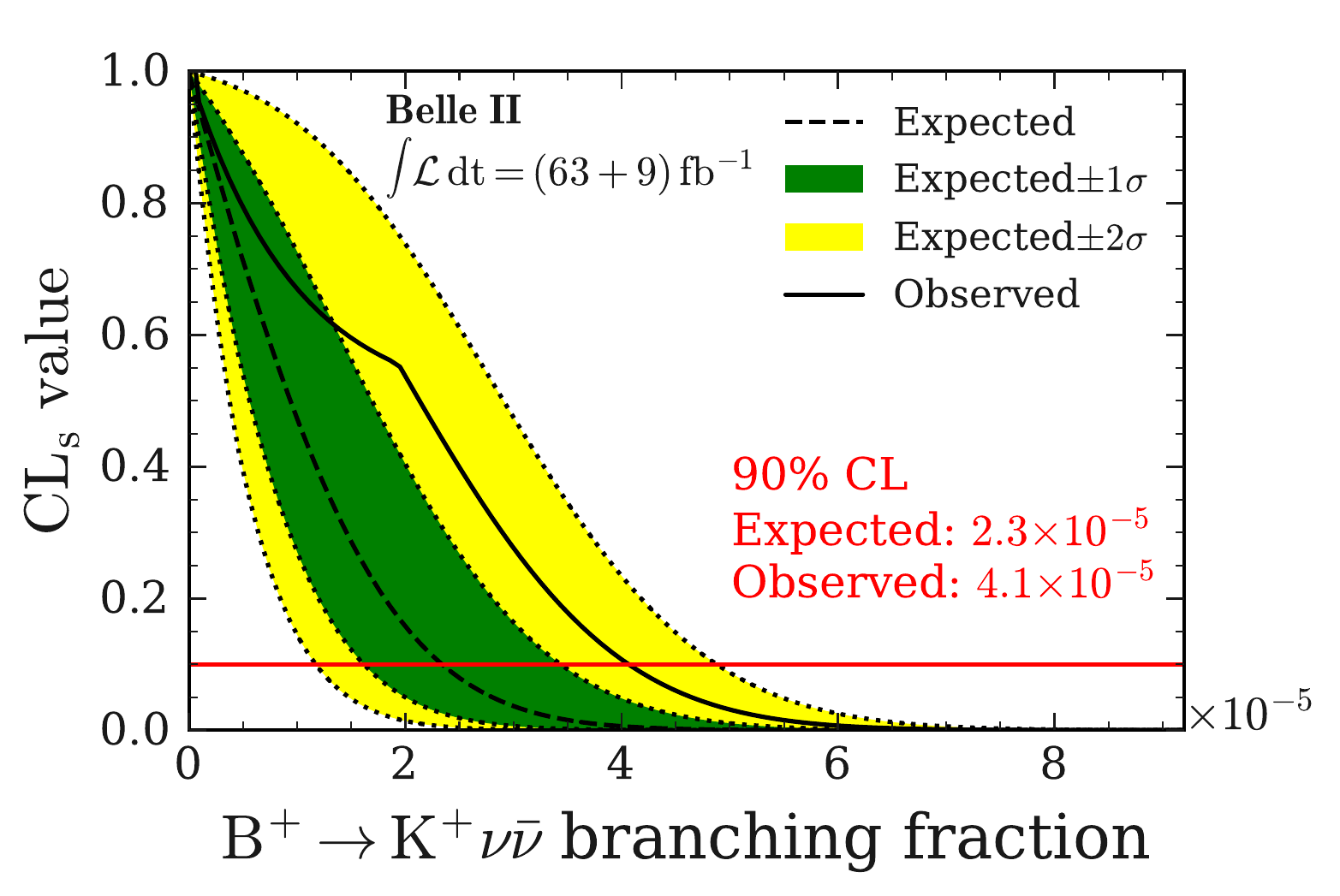}
\caption{
\CLs value as a function of the branching fraction of \BKnn for expected and observed signal yields and the corresponding upper limits at 90\% confidence level (CL).
The expected limit is derived for the background-only hypothesis.
The observed limit is derived from a simultaneous fit to the on-resonance and off-resonance data, corresponding to an integrated luminosity of 63\invfb and 9\invfb, respectively.
}
\end{figure}

\begin{figure}[htp]
\centering
\includegraphics[width=\linewidth]{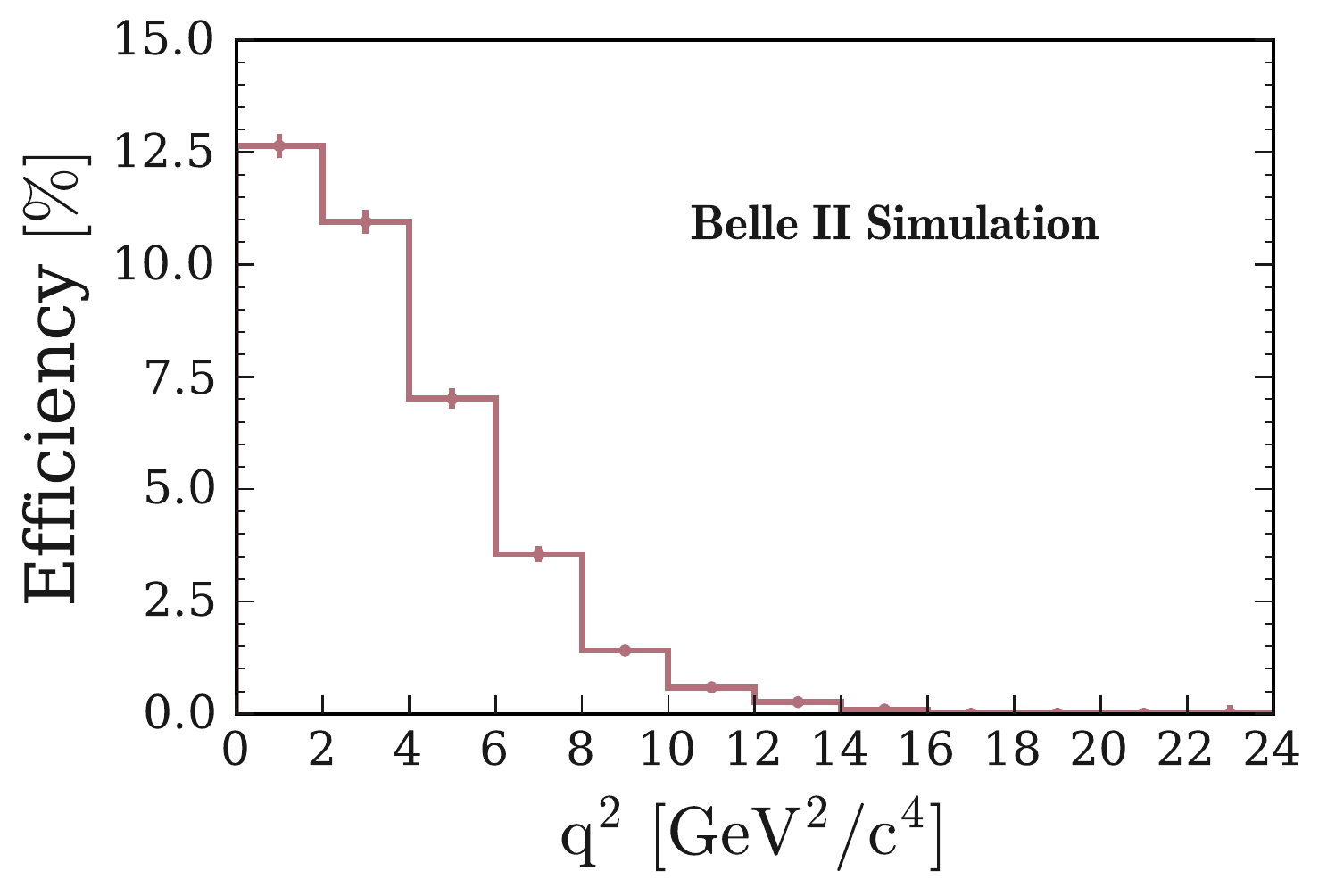}
\caption{Signal efficiency as a function of the dineutrino invariant mass squared $q^2$ for events in the SR ($\BDT1>0.9$ and $\BDT2>0.95$). The error bars indicate the statistical uncertainty.}
\end{figure}

\begin{figure}[htp]
\centering
\includegraphics[width=\linewidth]{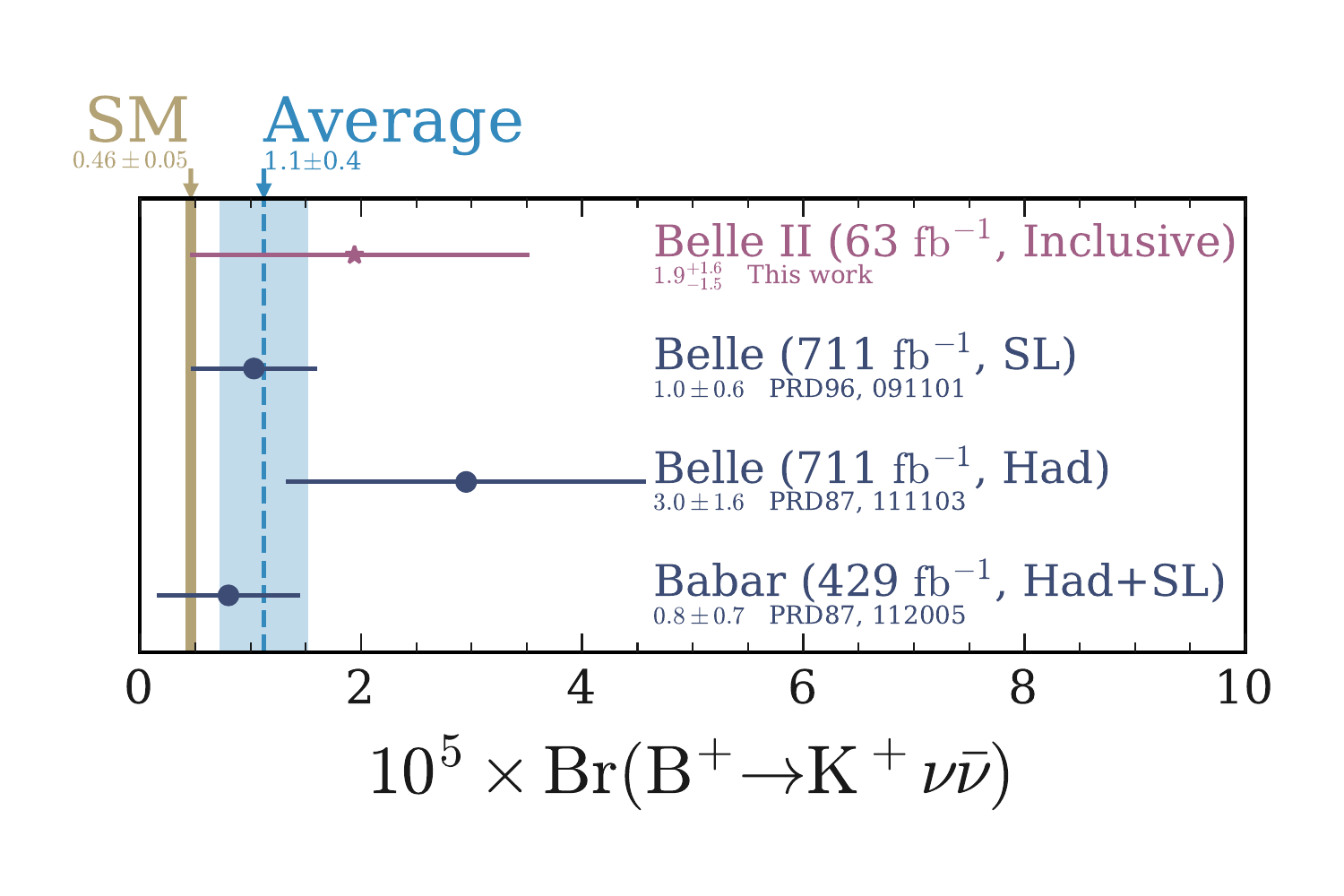}
\caption{
  The branching fraction measured by Belle II, previous experiments \cite{PhysRevD.87.111103,PhysRevD.87.112005,PhysRevD.96.091101} and the Standard Model expectation \cite{Blake:2016olu}.
  The values reported for Belle are computed based on the quoted observed number of events and efficiency.
  The weighted average is computed assuming that uncertainties are uncorrelated.
}
\end{figure}

\newpage

\bibliography{references}